%% file: 0KDD2018.tex
\documentclass[sigconf]{acmart}
\usepackage{booktabs} % For formal tables
\usepackage{amsmath}
\usepackage{amsfonts}
\usepackage{comment}
\usepackage{footmisc}
\usepackage{mathrsfs}
\usepackage{amssymb}
\usepackage{graphicx}
\usepackage{subfigure}
\usepackage{multirow}
\usepackage{algorithm}
\usepackage{algorithmic}
\usepackage{multicol}  
\usepackage{enumitem}
\usepackage{array}
\usepackage{float}
\usepackage{hyperref}

\begin{document}
\copyrightyear{2018} 
\acmYear{2018} 
\setcopyright{acmcopyright}
\acmConference[KDD '18]{The 24th ACM SIGKDD International Conference on Knowledge Discovery \& Data Mining}{August 19--23, 2018}{London, United Kingdom}
\acmBooktitle{KDD '18: The 24th ACM SIGKDD International Conference on Knowledge Discovery \& Data Mining, August 19--23, 2018, London, United Kingdom}
\acmPrice{15.00}
%\acmDOI{10.1145/3219819.3219886}
%\acmISBN{978-1-4503-5552-0/18/08}
%\fancyhead{}

\title{Recommendations with Negative Feedback via\\Pairwise Deep Reinforcement Learning}

\author{Xiangyu Zhao}
\affiliation{
	\institution{Data Science and Engineering Lab\\
		Michigan State University}
}
\email{zhaoxi35@msu.edu}

\author{Liang Zhang}
\affiliation{
	\institution{Data Science Lab\\Intelligent Advertising Lab\\
		JD.com}
}
\email{zhangliang16@jd.com}

\author{Zhuoye Ding}
\affiliation{
	\institution{Data Science Lab\\
		JD.com}
}
\email{dingzhuoye@jd.com}

\author{Long Xia}
\affiliation{
	\institution{Data Science Lab\\
		JD.com}
}
\email{xialong@jd.com}

\author{Jiliang Tang}
\affiliation{
	\institution{Data Science and Engineering Lab\\
		Michigan State University}
}
\email{tangjili@msu.edu}

\author{Dawei Yin}
\affiliation{
	\institution{Data Science Lab\\
		JD.com}
}
\email{yindawei@acm.org}

\renewcommand{\shortauthors}{Xiangyu Zhao et al.}

\begin{abstract}
	Recommender systems play a crucial role in mitigating the problem of information overload by suggesting users' personalized items or services. The vast majority of traditional recommender systems consider the recommendation procedure as a static process and make recommendations following a fixed strategy. In this paper, we propose a novel recommender system with the capability of continuously improving its strategies during the interactions with users. We model the sequential interactions between users and a recommender system as a Markov Decision Process (MDP) and leverage Reinforcement Learning (RL) to automatically learn the optimal strategies via recommending trial-and-error items and receiving reinforcements of these items from users' feedback.  Users' feedback can be positive and negative and both types of feedback have great potentials to boost  recommendations. However, the number of negative feedback is much larger than that of positive one; thus incorporating them  simultaneously is challenging since positive feedback could be buried by negative one.  In this paper, we develop a novel approach to incorporate them into the proposed deep recommender system (DEERS) framework. The experimental results based on real-world e-commerce data demonstrate the effectiveness of the proposed framework. Further experiments have been conducted to understand the importance of both positive and negative feedback in recommendations.
\end{abstract}


\maketitle

\input{1Introduction}
\input{2ProblemStatement}
\input{3Framework}
\input{4Experiments}
\input{5RelatedWork_Conclusion}

\section*{Acknowledgements}
This material is based upon work supported by, or in part by, the National Science Foundation (NSF) under grant number IIS-1714741 and IIS-1715940. 

\bibliographystyle{ACM-Reference-Format}
\bibliography{6Reference} 
\end{document}

%% file: 1Introduction.tex
\section{Introduction}
\label{sec:introduction}
Recommender systems are intelligent E-commerce applications. They assist users in their information-seeking tasks by suggesting items (products, services, or information) that best fit their needs and preferences. Recommender systems have become increasingly popular in recent years, and have been utilized in a variety of domains including movies, music, books, search queries, and social tags~\cite{resnick1997recommender,ricci2011introduction}. Typically, a recommendation procedure can be modeled as interactions between users and recommender agent (RA). It consists of two phases: 1) user model construction and 2) recommendation generation~\cite{mahmood2007learning}. During the interaction, the recommender agent builds a user model to learn users' preferences based on users' information or feedback. Then, the recommender agent generates a list of items that best match users' preferences.

Most existing recommender systems including collaborative filtering, content-based and learning-to-rank consider the recommendation procedure as a static process and make recommendations following a fixed greedy strategy. However, these approaches may fail given the dynamic nature of the users' preferences. Furthermore, the majority of existing recommender systems are designed to maximize the immediate (short-term) reward of recommendations, i.e., to make users purchase the recommended items, while completely overlooking whether these recommended items will lead to more profitable (long-term) rewards in the future~\cite{shani2005mdp}. 

In this paper, we consider the recommendation procedure as sequential interactions between users and recommender agent; and leverage Reinforcement Learning (RL) to automatically learn the optimal recommendation strategies. Recommender systems based on reinforcement learning have two advantages. First, they are able to continuously update their try-and-error strategies during the interactions, until the system converges to the optimal strategy that generates recommendations best fitting their users' dynamic preferences. Second, the models in the system are trained via estimating the present value with delayed rewards under current states and actions. The optimal strategy is made by maximizing the expected long-term cumulative reward from users. Therefore the system could identify items with small immediate rewards but making big contributions to the rewards for future recommendations.

Efforts have been made on utilizing reinforcement learning for recommender systems~\cite{shani2005mdp,taghipour2008hybrid}. For instance, the work~\cite{shani2005mdp} modeled the recommender system as a MDP process and estimated the transition probability and then the Q-value table. However, these methods may become inflexible with the increasing number of items for recommendations. This prevents them from being adopted in e-commence recommender systems. Thus, we leverage Deep Q-Network (DQN), an (adapted) artificial neural network, as a non-linear approximator to estimate the action-value function in RL. This model-free reinforcement learning method does not estimate the transition probability and not store the Q-value table. This makes it flexible to support huge amount of items in recommender systems. It can also enrich the system's generalization compared to traditional approaches that estimate action-value function separately for each sequence.

\begin{figure}
	\centering
	\includegraphics[width=63mm]{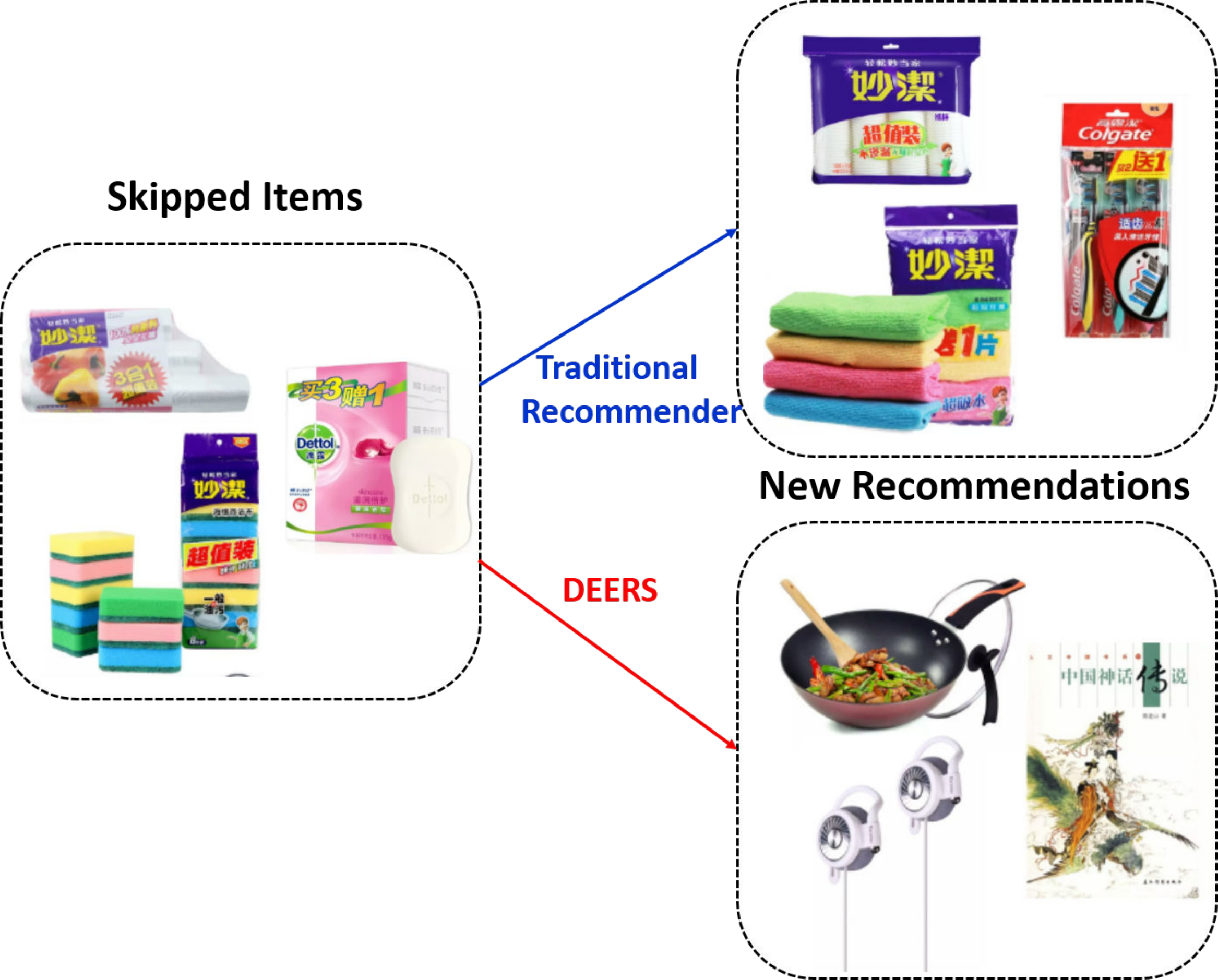}
	\caption{Impact of negative feedback on recommendations.}
	\label{fig:negative-samples}
\end{figure}

When we design recommender systems, positive feedback(clicked /ordered feedback) represents the users' preference and thus is the most important information to make recommendations. In reality, users also skip some recommended items during the recommendation procedure. These skipped items influence user's click/order behaviors~\cite{dupret2008user}, which can help us gain better understandings about users' preferences. Hence, it is necessary to incorporate such negative feedback. However, the number of skipped items (or negative feedback) is typically far larger than that of positive ones. Hence, it is challenging to capture both positive and negative feedback since positive feedback could be buried by negative one. In this paper, we propose a framework DEERS to model positive and negative feedback simultaneously. As shown in Figure~\ref{fig:negative-samples}, when items are skipped by a user, traditional recommender systems do not update their strategies, and still recommend similar items; while the proposed framework can capture these negative feedback and update the strategy for recommendations. We summarize our major contributions as follows: 
\begin{itemize}
	\item We identify the importance of negative feedback from the recommendation procedure and provide a principled approach to capture it for recommendations;
	\item We propose a deep reinforcement learning based framework DEERS, which can automatically learn the optimal recommendation strategies by incorporating positive and negative feedback;
	\item We demonstrate the effectiveness of the proposed framework in real-world e-commerce data and validate the importance of negative feedback in accurate recommendations.
\end{itemize}

The rest of this paper is organized as follows. In Section 2, we formally define the problem of recommender system via reinforcement learning. In Section 3, we provide approaches to model the recommending procedure as sequential user-agent interactions and introduce details about employing deep reinforcement learning to automatically learn the optimal recommendation strategies via offline-training. Section 4 carries out experiments based on real-word offline users' behavior log and presents experimental results. Section 5 briefly reviews related work. Finally, Section 6 concludes this work and discusses our future work.

%% file: 2ProblemStatement.tex
\section{Problem Statement}
\label{sec:problem} 
We study the recommendation task in which a recommender agent (RA) interacts with environment $\mathcal{E}$ (or users) by sequentially choosing recommendation items over a sequence of time steps, so as to maximize its cumulative reward. We model this problem as a Markov Decision Process (MDP), which includes a sequence of states, actions and rewards. More formally, MDP consists of a tuple of five elements $(\mathcal{S}, \mathcal{A}, \mathcal{P}, \mathcal{R}, \gamma)$ as follows:
\begin{itemize}
	\item {\bf State space $\mathcal{S}$}: A state $s_t \in \mathcal{S}$ is defined as the browsing history of a user before time $t$. The items in $s_t$ are sorted in chronological order.  
	\item {\bf Action space $\mathcal{A}$}:  The action $a_t \in \mathcal{A}$ of RA is to recommend items to a user at time $t$. Without the loss of generality, we assume that the agent only recommends one item to the user each time. Note that it is straightforward to extend it with recommending multiple items.  
	\item {\bf Reward $\mathcal{R}$}: After the RA taking an action $a_t$ at the state $s_t$, i.e., recommending an item to a user, the user browses this item and provides her feedback. She can skip (not click), click, or order this item, and the RA receives immediate reward $r(s_t,a_t)$ according to the user's feedback.
	\item {\bf Transition probability $\mathcal{P}$}: Transition probability $p(s_{t+1}|s_t,a_t)$ defines the probability of state transition from $s_t$ to $s_{t+1}$ when RA takes action $a_t$. We assume that the MDP satisfies $p(s_{t+1}|s_t,a_t,...,s_1,a_1) = p(s_{t+1}|s_t,a_t)$.
	\item {\bf Discount factor $\gamma$}: $\gamma \in [0,1]$ defines the discount factor when we measure the present value of future reward. In particular, when $\gamma = 0$, RA only considers the immediate reward. In other words, when $\gamma = 1$, all future rewards can be counted fully into that of the current action.
\end{itemize}

Figure \ref{fig:interaction} illustrates the agent-user interactions in MDP. With the notations and definitions above, the problem of item recommendation can be formally defined as follows: \textit{Given the historical MDP, i.e., $(\mathcal{S}, \mathcal{A}, \mathcal{P}, \mathcal{R}, \gamma)$, the goal is to find a recommendation policy $\pi:\mathcal{S} \rightarrow \mathcal{A}$, which can  maximize the cumulative reward for the recommender system.}

%% file: 3Framework.tex
\begin{figure}[t]
	\centering
	\includegraphics[width=81mm]{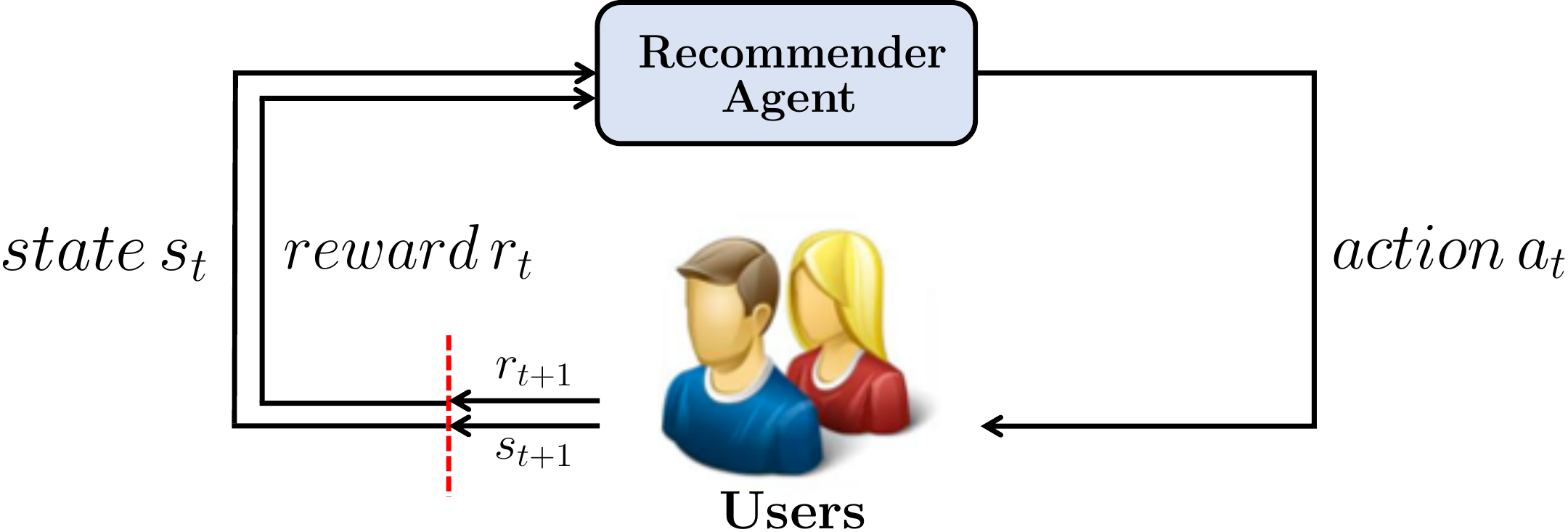}
	\caption{The agent-user interactions in MDP.}
	\label{fig:interaction}
\end{figure}
\section{The Proposed Framework based on Deep Reinforcement Learning with Negative Feedbacks}
\label{sec:framework}
Negative feedback dominates users' feedback to items and positive feedback could be buried by negative one if we aim to capture them simultaneously.  Thus in this section, we first propose a basic model that only considers positive feedback (clicked/ordered items) into the state. We build the deep reinforcement learning framework under this setting. After that, we consider negative feedback (skipped items) in the state space and redesign the deep architecture. Finally, we discuss how to train the framework via offline users' behavior log and how to utilize the framework for item recommendations. 

\subsection{The Basic DQN Model}
\label{sec:traditional}
The positive items represent the key information about users' preference, i.e., which items the users prefer to. A good recommender system should recommend the items users prefer the most. Thus, we only consider the positive feedback into the state in the basic model. More formally, we redefine the state and transition process for the basic model below:
\begin{itemize}
	\item {\bf State $s$}: State $s = \{i_{1}, \cdots, i_{N}\} \in \mathcal{S}$ is defined as the previous $N$ items that a user clicked/ordered recently. The items in state $s$ are sorted in chronological order. 
	\item {\bf Transition from $s$ to $s'$}: When RA recommends item $a$ at state $s$ to a user, the user may skip, click, or order the item. If the user skips the item, then the next state $s' = s$; while if the user clicks/orders the item, then the next state $s' = \{i_{2}, \cdots, i_{N}, a\}$.
\end{itemize}

Note that in reality, using discrete indexes to denote items is not sufficient since we cannot know the relations between different items only from indexes. One common way is to use extra information to represent items. For instance, we can use the attribute information like brand, price, sale per month, etc. Instead of extra item information, our model uses the RA/user interaction information, i.e., users' browsing history. We treat each item as a word and the clicked items in one recommendation session as a sentence. Then, we can obtain dense and low-dimensional vector representations for items by word embedding\cite{levy2014neural}.

As shown is Figure \ref{fig:interaction}, by interacting with the environment (users), recommender agent takes actions (recommends items) to users in such a way that maximizes the expected return, which includes the delayed rewards. We follow the standard assumption that delayed rewards are discounted by a factor of $\gamma$ per time-step, and define the action-value function $Q(s, a)$ as the expected return based on state $s$ and action $a$. The optimal action-value function $Q^*(s, a)$, which has the maximum expected return achievable by the optimal policy, should follow the Bellman equation \cite{bellman2013dynamic} as:
\begin{equation}\label{equ:Q*sa}
    Q^{*}(s, a)=\mathbb{E}_{s'} \, \big[r+\gamma\max_{a'}Q^{*}(s', a')|s,a\big].
\end{equation}

In real recommender systems, the state and action spaces are enormous, thus estimating $Q^{*}(s, a)$ by using the Bellman equation for each state-action pair is infeasible. In addition, many state-action pairs may not appear in the real trace such that it is hard to update their values. Therefore, it is more flexible and practical to use an approximator function to estimate the action-value function, i.e., $Q^{*}(s,a) \approx Q(s, a; \theta)$ . In practice, the action-value function is usually highly nonlinear. Deep neural networks are known as excellent approximators for non-linear functions. In this paper, We refer to a neural network function approximator with parameters $\theta$ as a $\mathrm{Q}$-network. A $\mathrm{Q}$-network can be trained by minimizing a sequence of loss functions $L(\theta)$ as:
\begin{equation}\label{equ:L}
L(\theta)=\mathbb{E}_{s, a,r,s'}\big[(y-Q(s, a;\theta))^{2}\big],
\end{equation}
where $y= \mathbb{E}_{s'}[r+\gamma\max_{a'}Q(s',\ a';\theta^{p})|s, a]$ is the target for the current iteration. The parameters from the previous iteration $\theta^{p}$ are fixed when optimizing the loss function $L(\theta)$. The derivatives of loss function $L(\theta)$ with respective to parameters $\theta$ are presented as follows:
\begin{equation}\label{equ:differentiating}
\begin{aligned}
\nabla_{\theta}L(\theta) & =\mathbb{E}_{s, a, r, s'}\big[(r+\gamma\max_{a'}Q(s',a';\theta^{p})\\
&-Q(s, a;\theta))\nabla_{\theta}Q(s, a;\theta)\big].
\end{aligned}
\end{equation}

\begin{figure}[t]
	\centering
	\includegraphics[width=63mm]{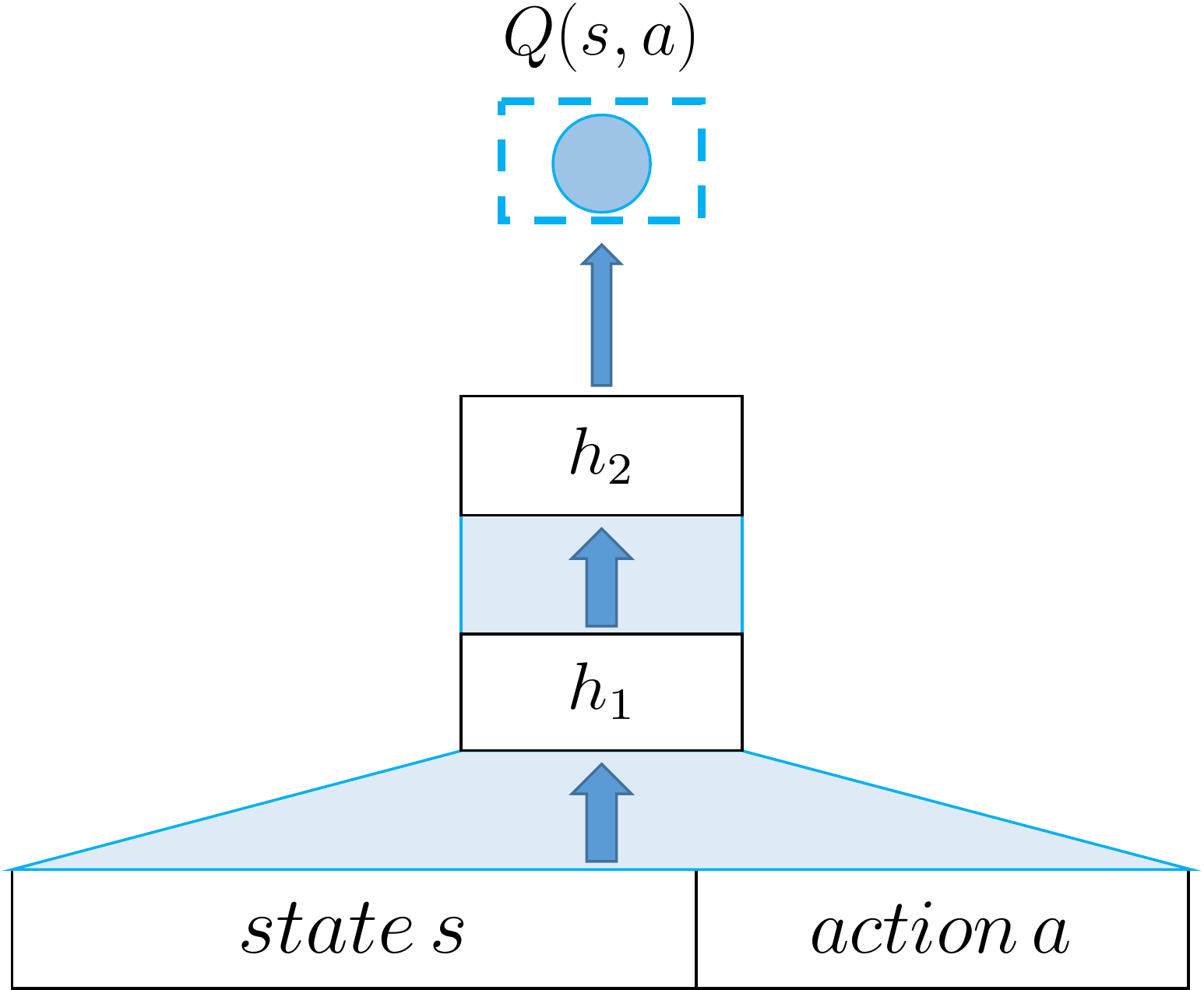}
	\caption{The architecture of the basic DQN model for recommendations.}
	\label{fig:DQN1}
\end{figure}

In practice, it is often computationally efficient to optimize the loss function by stochastic gradient descent, rather than computing the full expectations in the above gradient. Figure \ref{fig:DQN1} illustrates the architecture of basic DQN model.

\subsection{The Proposed DEERS Framework}
\label{sec:nagetive}
As previously discussed, the items that users skipped can also indicate users' preferences, i.e., what users may not like. However, the system with only positive items will not change its state or update its strategy when users skip the recommended items. Thus, the state should not only contain positive items that user clicked or ordered, but also incorporate negative (skipped) items. To integrate negative items into the model, we redefine state and transition process as follows:
\begin{itemize}
\item {\bf State $s$}: $s = (s_{+}, s_{-}) \in S$ is a state, where $s_{+} = \{i_1, \cdots, i_{N}\}$ is defined as the previous $N$ items that a user clicked or ordered recently, and $s_{-} = \{j_1, \cdots, j_{N}\}$ is the previous $N$ items that the user skipped recently. The items in $s_{+}$ and $s_{-}$ are processed in chronological order.
\item {\bf Transition from $s$ to $s'$}: When RA recommends item $a$ at state $s = (s_{+}, s_{-})$ to a user, if users skip the recommended item, we keep $s'_+ = s_{+}$ and update $s'_- = \{j_2, \cdots, j_{N}, a\}$. If users click or order the recommended item, update $s'_+ = \{i_2, \cdots, i_{N}, a\}$, while keeping $s'_- = s_{-}$. Then set $s' = (s'_{+}, s'_{-})$. 
\end{itemize}

We can see from the new definition that no matter users accept or skip recommendations, the new system will incorporate the feedback for next recommendations. Also it can distinguish positive and negative feedback with $s_{+}$ and $s_{-}$. 

\subsubsection{\textbf{The Architecture of DEERS Framework}}
\begin{figure}[t]
	\centering
	\includegraphics[width=72mm]{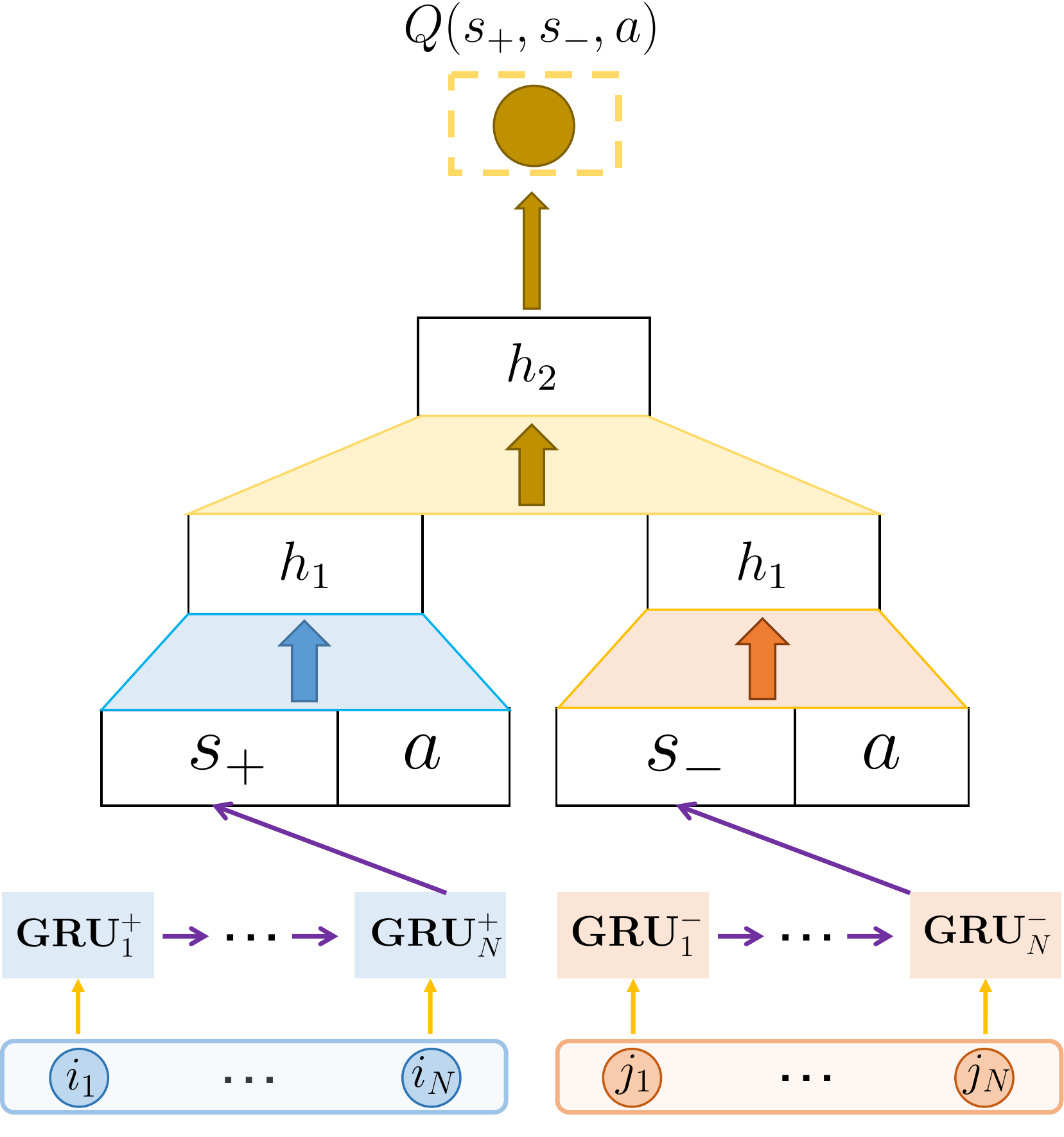}
	\caption{The architecture of the DEERS framework.}
	\label{fig:DQN2}
\end{figure}

In this approach, we concatenate positive state $s_{+}$ and a recommended item $a$ as positive input (positive signals), while concatenating negative state $s_{-}$ and the recommended item $a$ as the negative input (negative signals). Rather than following the basic reinforcement learning architecture in Figure \ref{fig:DQN1}, we construct DQN in a novel way. Figure \ref{fig:DQN2} illustrates our new DQN architecture. Instead of just concatenating clicked/ordered items $\{i_1, \cdots, i_{N}\}$ as positive state $s_{+}$, we introduce a RNN with Gated Recurrent Units (GRU) to capture users' sequential preference as positive state $s_{+}$:
\begin{equation}\label{equ:update}
z_n=\sigma(W_{z}i_{n}+U_{z}h_{n-1}),
\end{equation}
\begin{equation}\label{equ:reset}
r_n=\sigma(W_{r}i_{n}+U_{r}h_{n-1}),
\end{equation}
\begin{equation}\label{equ:gru1}
\hat{h}_n=\tanh[W i_{n}+U (r_n\cdot h_{n-1})],
\end{equation}
\begin{equation}\label{equ:gru}
h_{n}=(1-z_{n})h_{n-1}+z_n \hat{h}_n,
\end{equation}
where GRU utilizes an update gate $z_n$ to generate a new state and a reset gate $r_n$ to control the input of the former state $h_{n-1}$. The inputs of GRU are the embeddings of user's recently clicked /ordered items $\{i_1, \cdots, i_{N}\}$, while we use the output (final hidden state $h_{N}$) as the representation of the positive state, i.e., $s_{+}=h_{N}$. We obtain negative state $s_{-}$ following a similar way. Here we leverage GRU rather than Long Short-Term Memory (LSTM) because that GRU outperforms LSTM for capturing users' sequential behaviors in recommendation task \cite{hidasi2015session}.

As shown in Figure \ref{fig:DQN2}, we feed positive input (positive signals) and negative input (negative signals) separately in the input layer. Also, different from traditional fully connected layers, we separate the first few hidden layers for positive input and negative input. The intuition behind this architecture is to recommend an item that is similar to the clicked/ordered items (left part), while dissimilar to the skipped items (right part). This architecture can assist DQN to capture distinct contributions of the positive and negative feedback to recommendations.

Then, the loss function $L(\theta)$ for an iteration of the new deep $\mathrm{Q}$-network training can be rewritten as:
\begin{equation}\label{equ:L1}
L(\theta)=\mathbb{E}_{s, a,r,s'}\bigg[\bigg(y-Q(s_{+}, s_{-}, a;\theta)\bigg)^{2}\bigg],
\end{equation}
where $y= \mathbb{E}_{s'}[r+\gamma\max_{a'}Q(s'_{+}, s'_{-},a';\theta^{p})|s_{+}, s_{-}, a]$ is the target of current iteration. Then, the gradient of loss function with respect to the $\theta$ can be computed as:
\begin{equation}\label{equ:differentiating1}
\begin{aligned}
\nabla_{\theta}L(\theta) & =\mathbb{E}_{s, a, r, s'}\bigg[\bigg(r+\gamma\max_{a'}Q(s'_{+}, s'_{-},a';\theta^{p})\\
&-Q(s_{+}, s_{-}, a;\theta)\bigg)\nabla_{\theta}Q(s_{+}, s_{-}, a;\theta)\bigg].
\end{aligned}
\end{equation}

To decrease the computational cost of Equation (\ref{equ:differentiating1}), we adopt item recalling mechanism to reduce the number of relevant items~\footnote{ In general, user's current preference should be related to user's previous clicked/ordered items  (say $\mathcal{L}_p$). Thus for each item in $\mathcal{L}_p$, we collect a number of most similar items from the whole item space, and combine all collected items as the item space for current recommendation session.}.

\subsubsection{\textbf{The Pairwise Regularization Term}}
With deep investigations on the users' logs, we found that in most recommendation sessions, the RA recommends some items belong to the same category (e.g. telephone), while users click/order a part of them and skip others. We illustrate a real example of a recommendation session in Table \ref{table:overall result}, in which three categories of items are recommended to user, and each time the RA recommends one item to the user. For category B, we can observe that the user clicked item $a_2$ while skipped item $a_5$, which indicates the partial order of user's preference over these two items in category B. This partial order naturally inspires us maximizing the difference of Q-values between $Q(s_2, a_2)$ and $Q(s_5, a_5)$. At time 2, we name $a_5$ as the \textit{competitor item} of  $a_2$. Sometimes, one item could have multiple ``competitor'' items; thus we select the item at the closest time as the ``competitor'' item. For example, at time 3, $a_3$'s competitor item is $a_1$ rather than $a_6$. Overall, we select one target item's  \textit{competitor item} according to three requirements: 1) the ``competitor'' item belongs to the same category with the target item; 2) user gives different types of feedback to the ``competitor'' item and the target item; and 3)  the ``competitor'' item is at the closest time to the target item.

\begin{table}[t]
	\Large
	\caption{An illustrative example of a recommendation session.}
	\begin{tabular}{ccccc}
		\hline
		Time & State & Item & Category & Feedback \\ \hline
		1     & $s_1$     & $a_1$    & A        & skip        \\
		2     & $s_2$    & $a_2$    & B        & click        \\
		3     & $s_3$     & $a_3$    & A        & click       \\
		4     & $s_4$     & $a_4$    & C        & skip       \\
		5     & $s_5$     & $a_5$    & B        &  skip       \\
		6     & $s_6$     & $a_6$    & A        & skip        \\
		7     & $s_7$     & $a_7$    & C        & order        \\ \hline
	\end{tabular}
	\label{table:overall result}
\end{table}

To maximize the difference of Q-values between target and competitor items, we add a regularization term to Equation \ref{equ:L1}: 
\begin{equation}\label{equ:L1_1}
\begin{aligned}
L(\theta)&=\mathbb{E}_{s, a,r,s'}\bigg[\bigg(y-Q(s_{+}, s_{-}, a;\theta)\bigg)^{2}\\
&-\alpha \bigg(Q(s_{+}, s_{-}, a;\theta) - Q(s_{+}, s_{-}, a^C;\theta)\bigg)^2\bigg],
\end{aligned}
\end{equation}
where $y= \mathbb{E}_{s'}[r+\gamma\max_{a'}Q(s'_{+}, s'_{-},a';\theta^{p})|s_{+}, s_{-}, a]$ is the target for the iteration. The second term aims to maximize the difference of Q-values between target item $a$ and competitor item $a^C$ at state $(s_{+}, s_{-})$, which is controlled by a non-negative parameter $\alpha$. Note that since user's preference is relatively stable during a short time slot \cite{wu2017returning}, we assume that user will give same feedback to $a^C$ at state $(s_{+}, s_{-})$. For example,  if RA recommends item $a_5$ at state $s_2$, the user will still skip $a_5$. The gradient of loss function can be computed as:
\begin{equation}\label{equ:differentiating2}
\begin{aligned}
\nabla_{\theta}L(\theta) & =\mathbb{E}_{s, a, r, s'}\bigg[\bigg(y-Q(s_{+}, s_{-}, a;\theta)\bigg)\nabla_{\theta}Q(s_{+}, s_{-}, a;\theta)\\
&- \alpha \bigg(Q(s_{+}, s_{-}, a;\theta) - Q(s_{+}, s_{-}, a^C;\theta)\bigg)\\
&\bigg( \nabla_{\theta}Q(s_{+}, s_{-}, a;\theta) - \nabla_{\theta}Q(s_{+}, s_{-}, a^C;\theta)\bigg)\bigg].
\end{aligned}
\end{equation}

\subsection{Off-policy Training Task}
\label{sec:offline}
With the proposed deep reinforcement learning framework, we can train the parameters in models and then do the test work. We train the proposed model based on users' offline log, which records the interaction history between RA's policy $b(s_t)$ and users' feedback. RA takes the action based on the off-policy $b(s_t)$ and obtain the feedback from the offline log. We present our off-policy training algorithm in details shown in Algorithm \ref{alg:model1}. Note that our algorithm is \textit{model-free}: it solves the reinforcement learning task directly using samples from the environment, without explicitly constructing an estimate of the environment.

\begin{algorithm}[t]
	\caption{\label{alg:model1} Off-policy Training of DEERS Framework.}
	\raggedright
	\begin{algorithmic} [1]
		\STATE Initialize the capacity of replay memory $\mathcal{D}$
		\STATE Initialize action-value function $Q$ with random weights
		\FOR{session $=1, M$}
		\STATE  Initialize state $s_{0}$ from previous sessions
		\FOR{$t=1, T$}
        \STATE  Observe state $s_t=\{i_1, \cdots, i_{N},j_1, \cdots, j_{N}\}$ 
		\STATE  Execute action $a_{t}$ following off-policy $b(s_t)$ 
        \STATE  Observe reward $r_{t}$ from users
		\STATE  Set $s_{t+1}=
		\left\{\begin{array}{ll}
		\{i_2, \cdots, i_{N},a_t, j_1, \cdots, j_{N}\}
		& r_{t} > 0\\
		\{i_1, \cdots, i_{N}, j_2, \cdots, j_{N}, a_t\} 
		& r_{t} = 0
		\end{array}\right.$ 
		\STATE  Find competitor item $a_t^C$ of $a_{t}$
		\STATE  Store transition $(s_{t}, a_{t},  r_{t}, s_{t+1},a_t^C/null)$ in $\mathcal{D}$
		\STATE  Sample minibatch of transitions $(s, a, r, s',a^C/null)$ from $\mathcal{D}$
		\STATE  Set $y=
		\left\{\begin{array}{ll}
		r 
		& \mathrm{terminal\,}  s'\\
		r+\gamma\max_{a'}Q(s',a';\theta) 
		& \mathrm{non-terminal\,} s'
		\end{array}\right.$
		\IF{$a^C$ exists} 
		\STATE Minimize $\big(y-Q(s, a;\theta)\big)^{2}-\alpha\big(Q(s, a;\theta)-Q(s, a^C;\theta)\big)^{2}$ according to Equ (\ref{equ:differentiating2})
		\ELSE
		\STATE Minimize  $\big(y-Q(s, a;\theta)\big)^{2}$ according to Equ (\ref{equ:differentiating1})
		\ENDIF
		\ENDFOR
		\ENDFOR
	\end{algorithmic}
\end{algorithm}

In each iteration on a training recommendation session, there are two stages. For storing transitions stage: given the state $s_t$, the recommender agent first recommends an item $a_t$ from a fixed off-policy $b(s_t)$ (line 7), which follows a standard off-policy way \cite{degris2012off}; then the agent observes the reward $r_t$ from users (line 8) and updates the state to $s_{t+1}$ (line 9) and try to find the competitor item $a_t^C$ (line 10); and finally the recommender agent stores transitions $(s_t,a_t,r_t,s_{t+1},a_t^C/null)$ into replay memory $\mathcal{D}$ (line 11). For model training stage: the recommender agent samples minibatch of transitions $(s, a, r, s',$ $a^C/null)$ from replay memory $\mathcal{D}$ (line 12), and then updates the parameters according to Equation (\ref{equ:differentiating1}) or Equation (\ref{equ:differentiating2}) (lines 13-18).

In the algorithm, we introduce widely used techniques to train our framework. For example, we utilize a technique known as {\it experience replay} \cite{lin1993reinforcement} (lines 1,11,12), and introduce separated evaluation and target networks~\cite{mnih2013playing}, which can help smooth the learning and avoid the divergence of parameters. Moreover, we leverage {\it prioritized sampling strategy}~\cite{moore1993prioritized} to assist the framework learning from the most important historical transitions. 

\subsection{Offline Test}
\label{sec:test}
In the previous subsection, we have discussed how to train a DQN-based recommender model. Now we formally present the offline test of our proposed framework DEERS. 

The intuition of our offline test method is that, for a given recommendation session, the recommender agent reranks the items in this session according to the items' Q-value calculated by the trained DQN, and recommends the item with maximal Q-value to the user. Then the recommender agent observes the reward from users and updates the state. The reason why recommender agent only reranks items in this session rather than items in the whole item space is that for the historical offline dataset, we only have the ground truth rewards of the existing items in this session. The offline test algorithm in details is presented in Algorithm \ref{alg:test}.
\begin{algorithm}
	\caption{\label{alg:test} Offline Test of DEERS Framework.}
	\raggedright
    {\bf Input}: Initial state $s_{1}$, items ${ l_1, \cdots, l_T}$ and corresponding rewards ${r_1, \cdots, r_T}$  of a session $\mathcal{L}$.\\
    {\bf Output}:Recommendation list with new order\\
	\begin{algorithmic} [1]
		\STATE \textbf{for} $t =1, T$ \textbf{do}
		\STATE \quad Observe state $s_t = \{i_1, \cdots, i_{N}, j_1, \cdots, j_{N}\}$
        \STATE \quad Calculate Q-values of items in $\mathcal{L}$
        \STATE \quad Recommend item $l_{max}$ with maximal Q-value
        \STATE \quad Observe reward $r_t$ from users (historical logs)
		\STATE \quad Set $s_{t+1}=
		\left\{\begin{array}{ll}
		\{i_2, \cdots, i_{N},l_{max}, j_1, \cdots, j_{N}\}
		& r > 0\\
		\{i_1, \cdots, i_{N}, j_2, \cdots, j_{N}, l_{max}\} 
		& r = 0
		\end{array}\right.$ 
		\STATE \quad Remove $l_{max}$ from $\mathcal{L}$
		\STATE \textbf{end for}
	\end{algorithmic}
\end{algorithm}

In each iteration of a test recommendation session $\mathcal{L}$, given the state $s_t$ (line 2), the recommender agent recommends an item $l_{max}$ with maximal Q-value calculated by the trained DQN (lines 3-4), then observes the reward $r_t$ from users (line 5) and updates the state to $s_{t+1}$ (line 6), and finally it removes item $l_{max}$ from $\mathcal{L}$ (line 7). Without the loss of generality, the RA recommends one item to the user each time, while it is straightforward to extend it by recommending multiple items.

\subsection{Online Test}
\label{sec:on_testing}
We also do online test on a simulated online environment. The simulated online environment is also trained on users' logs, but not on the same data for training the DEERS framework. The simulator has the similar architecture with DEERS, while the output layer is a softmax layer that predicts the immediate feedback according to current state $s_t$ and a recommended item $a_t$. We test the simulator on users' logs(not the data for training the DEERS framework and simulator), and experimental results demonstrate that the simulated online environment has overall 90\% precision for  immediate feedback prediction task. This result suggests that the simulator can accurately simulate the real online environment and predict the online rewards, which enables us to test our model on it. 

When test DEERS framework on the well trained simulator, we can feed current state $s_t$ and a recommended item $a_t$ into the simulator, and receive immediate reward $r(s_t,a_t)$ from it.  The online test algorithm in details is presented in Algorithm \ref{alg:test_on}. Note that we can control the length of the session (the value $T$) manually.
\begin{algorithm}
	\caption{\label{alg:test_on}Online Test of DEERS Framework.}
	\raggedright
	\begin{algorithmic} [1]
		\STATE Initialize action-value function $Q$ with well trained weights
		\STATE \textbf{for} session $=1, M$ \textbf{do}
		\STATE \quad Initialize state $s_{1}$ from previous sessions
		\STATE \quad \textbf{for} $t=1, T$ \textbf{do}
		\STATE \qquad Observe state $s_t=\{i_1, \cdots, i_{N},j_1, \cdots, j_{N}\}$ 
		\STATE \qquad Execute action $a_{t}$ following policy $\pi$ 
		\STATE \qquad Observe reward $r_{t}$ from users (online simulator)
		\STATE \qquad Set $s_{t+1}=
		\left\{\begin{array}{ll}
		\{i_2, \cdots, i_{N},a_t, j_1, \cdots, j_{N}\}
		& r_{t} > 0\\
		\{i_1, \cdots, i_{N}, j_2, \cdots, j_{N}, a_t\} 
		& r_{t} = 0
		\end{array}\right.$ 
		\STATE \quad \textbf{end for}
		\STATE \textbf{end for}
	\end{algorithmic}
\end{algorithm}

In each iteration of a test recommendation session $\mathcal{L}$, given the state $s_t$ (line 5), the RA recommends an item $a_t$ following  policy $\pi$ (line 6), which is obtained in model training stage. Then the RA feed $s_t$ and $a_t$ into the simulated online environment and observes the reward $r_t$ (line 7) and updates the state to $s_{t+1}$ (line 8). 

%% file: 4Experiments.tex
\section{Experiments}
\label{sec:experiments}
In this section, we conduct extensive experiments with a dataset from a real e-commerce site to evaluate the effectiveness of the proposed framework. We mainly focus on two questions: (1) how the proposed framework performs compared to representative baselines; and (2) how the negative feedback (skipped items) contribute to the performance. We first introduce experimental settings. Then we seek answers to the above two questions. Finally, we study the impact of important parameters on the performance of the proposed framework. 

\subsection{Experimental Settings}
\label{sec:experimental_settings}
We evaluate our method on a dataset of September, 2017 from JD.com. We collect 1,000,000 recommendation sessions (9,136,976 items) in temporal order, and use first 70\% as training set and other 30\% as test set. For a given session, the initial state is collected from the previous sessions of the user. In this paper, we leverage $N = 10$ previously clicked/ordered items as positive state and $N = 10$ previously skipped items as negative state. The reward $r$ of skipped/clicked/ordered items are empirically set as 0, 1, and 5, respectively. The dimension of the embedding of items is 50, and we set the discounted factor $\gamma = 0.95$. For the parameters of the proposed framework such as $\alpha$ and $\gamma$, we select them via cross-validation. Correspondingly, we also do parameter-tuning for baselines for a fair comparison. We will discuss more details about parameter selection for the proposed framework in the following subsections. 

For the architecture of Deep Q-network, we leverage a 5-layer neural network, in which the first 3 layers are separated for positive and negative signals, and the last 2 layers connects both positive and negative signals, and outputs the Q-value of a given state and action.

As we consider our offline test task as a reranking task, we select \textbf{MAP}~\cite{turpin2006user} and \textbf{NDCG@40}~\cite{jarvelin2002cumulated} as the metrics to evaluate the performance.  The difference of ours from traditional Learn-to-Rank methods is that we rerank both clicked and ordered items together, and set them by different rewards, rather than only rerank clicked items as that in Learn-to-Rank problems. For online test, we leverage the accumulated rewards in the session as the metric.

\subsection{Performance Comparison for Offline Test}
\label{sec:ev_overall}
We compare the proposed framework with the following representative baseline methods: 
\begin{itemize}
	\item \textbf{CF}: Collaborative filtering\cite{breese1998empirical} is a method of making automatic predictions about the interests of a user by collecting preference information from many users, which is based on the hypothesis that people often get the best recommendations from someone with similar tastes to themselves. 
	\item \textbf{FM}: Factorization Machines\cite{rendle2010factorization} combine the advantages of support vector machines with factorization models. Compared with matrix factorization, higher order interactions can be modeled using the dimensionality parameter.
	\item \textbf{GRU}: This baseline utilizes the Gated Recurrent Units (GRU) to predict what user will click/order next based on the clicking/ordering histories. To make comparison fair, it also keeps previous $N = 10$ clicked/ordered items as states.
	\item \textbf{DEERS-p}: We use a Deep Q-network\cite{mnih2013playing} with embeddings of users' historical clicked/ordered items (state) and a recommended item (action) as input, and train this baseline following Eq.(\ref{equ:L}). Note that the state is also captured by GRU. 
\end{itemize}

\begin{figure}[t]
	\centering
	\includegraphics[width=81mm]{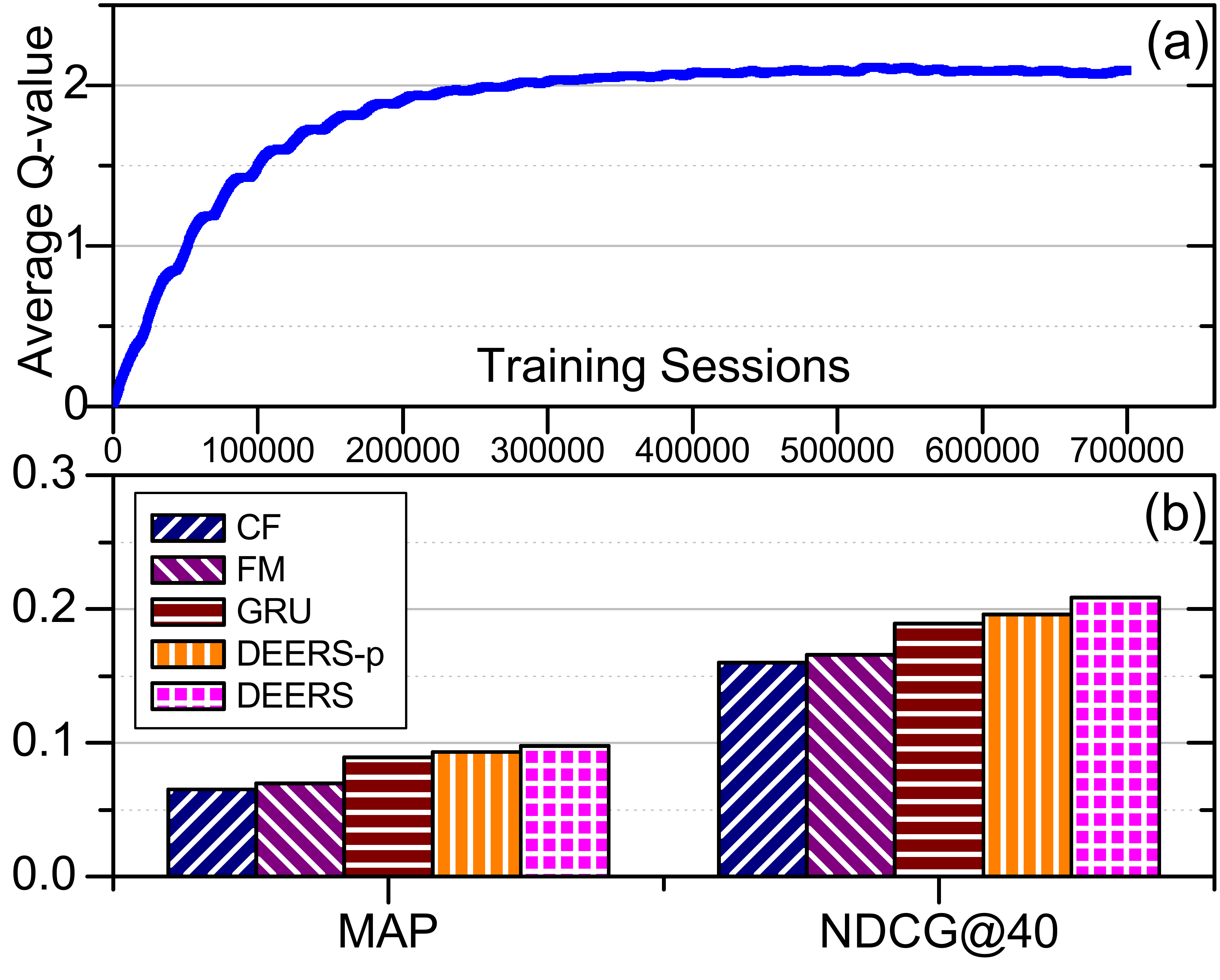}
	\caption{Overall performance comparison in offline test.}
	\label{fig:overall}
\end{figure}

The results are shown in Figure \ref{fig:overall}. We make following observations:
\begin{itemize}
	\item Figure \ref{fig:overall} (a) illustrates the training process of DEERS. We can observe that the framework approaches to convergence when the model are trained by 500,000 sessions. The fluctuation of the curve causes by the parameter replacement from evaluation DQN to target DQN.
	\item CF and FM achieve worse performance than GRU, DQN and DEERS, since CF and FM ignore the temporal sequence of the users' browsing history, while GRU can capture the temporal sequence, and DEERS-p and DEERS are able to continuously update their strategies during the interactions.
	\item GRU achieves worse performance than DEERS-p, since we design GRU to maximize the immediate reward for recommendations, while DEERS-p is designed to achieve the trade-off between short-term and long-term rewards. This result suggests that introducing reinforcement learning can improve the performance of recommendations.
	\item DEERS performs better than DEERS-p because we integrate both positive and negative items (or feedback) into DEERS, while DEERS-p is trained only based on positive items. This result indicates that negative feedback can also influence the decision making process of users, and integrating them into the model can lead to accurate recommendations. 
\end{itemize}

To sum up, we can draw answers to the two questions: (1) the proposed framework outperforms representative baselines; and (2) negative feedback can help the recommendation performance.

\subsection{Performance Comparison for Online Test}
We do online test on the aforementioned simulated online environment, and compare with DEERS with GRU and several variants of DEERS. Note that we do not include {\bf CF} and {\bf FM} baselines as offline test since CF and FM are not applicable to the simulated online environment.

\begin{itemize}
	\item \textbf{GRU}: The same GRU as in the above subsection. 	
	\item \textbf{DEERS-p}: The same DEERS-p as in the above subsection. 	
	\item \textbf{DEERS-f}: This variant is a traditional 5-layer DQN where all layers are fully connected. Note that the state is also captured by GRU. 
	\item \textbf{DEERS-t}: In this variant, we remove GRU units and just concatenate the previous $N = 10$ clicked/ordered items as positive state and previous $N = 10$ skipped items as negative state. 
	\item \textbf{DEERS-r}: This variant is to evaluate the performance of pairwise regularization term, so we set $\alpha = 0$ to eliminate the pairwise regularization term.
\end{itemize}
\begin{figure}[t]
	\centering
	\includegraphics[width=81mm]{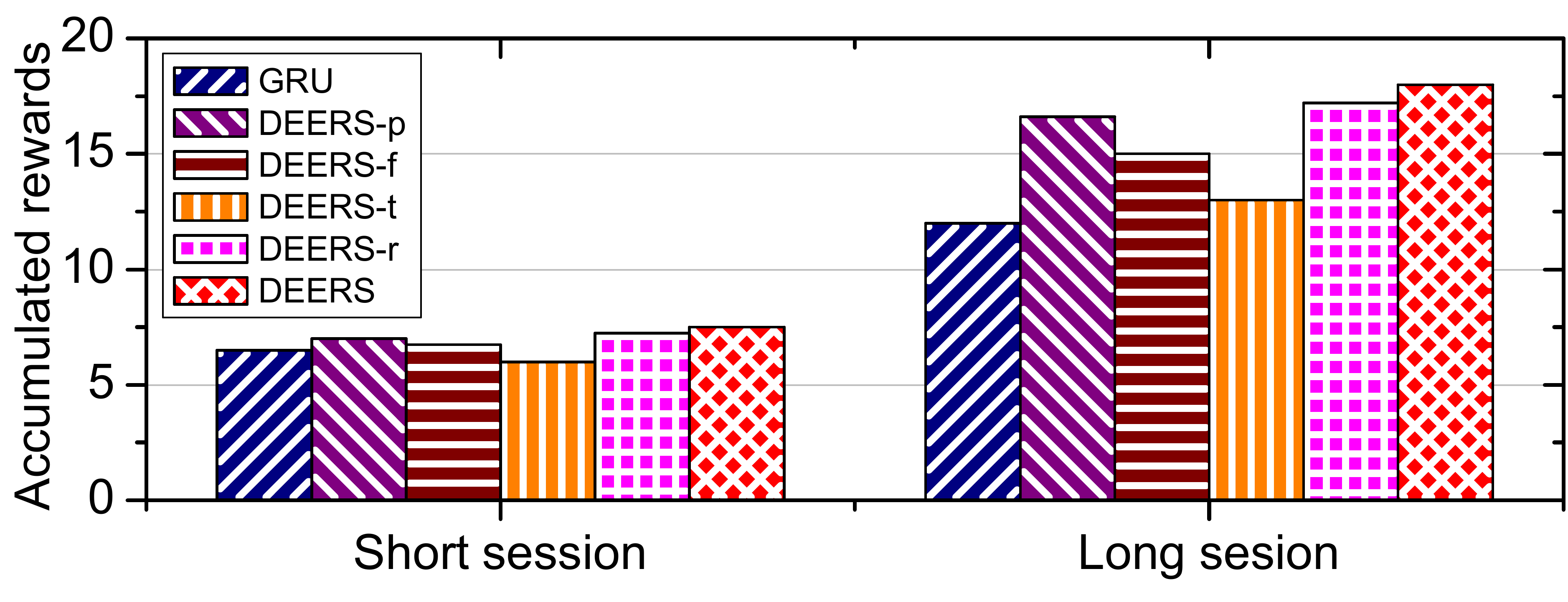}
	\caption{Overall performance comparison in online test.}
	\label{fig:architecture}
\end{figure}

As the test stage is based on the simulator, we can artificially control the length of recommendation sessions to study the performance in short and long sessions. We define short sessions with 100 recommended items, while long sessions with 300 recommended items. The results are shown in Figure \ref{fig:architecture}. It can be observed: 

\begin{itemize}
\item DEERS outperforms DEERS-f, which demonstrates that the new architecture of DQN can indeed improve the recommendation performance.
\item An interesting discovery is that DEERS-f achieves even worse performance than DEERS-p, which is trained only based on positive feedback. This result indicates that a full connected DQN cannot detect the differences between positive and negative items, and simply concatenating positive and negative feedback as input will reduce the performance of traditional DQN. Thus redesigning DQN architecture is a necessary.
\item In short recommendation sessions, GRU and DEERS-p achieve comparable performance. In other words, GRU and reinforcement learning models like DEERS-p can both recommend proper items matching users' short-term interests.
\item In long recommendation sessions, DEERS-p outperforms GRU significantly, because GRU is designed to maximize the immediate reward for recommendations, while reinforcement learning models like DEERS-p are designed to achieve the trade-off between short-term and long-term rewards. 
\item DEERS outperforms DEERS-t and DEERS-r, which suggests that introducing GRU to capture users' sequential preference and introducing pairwise regularization term can improve the performance of recommendations.
\end{itemize}

In summary, appropriately redesigning DQN architecture to incorporate negative feedback, leveraging GRU to capture users' dynamic preference and introducing pairwise regularization term can boost the recommendation performance. 

\subsection{Parameter Sensitivity Analysis}
\label{sec:parametric}
Our method has two key parameters, i.e., $\alpha$ that controls the pairwise regularization term and $N$ that controls the length of state. To study the impact of these parameters, we investigate how the proposed framework DEERS works with the changes of one parameter, while fixing other parameters.

\begin{figure}[t]
	\centering
	\includegraphics[width=81mm]{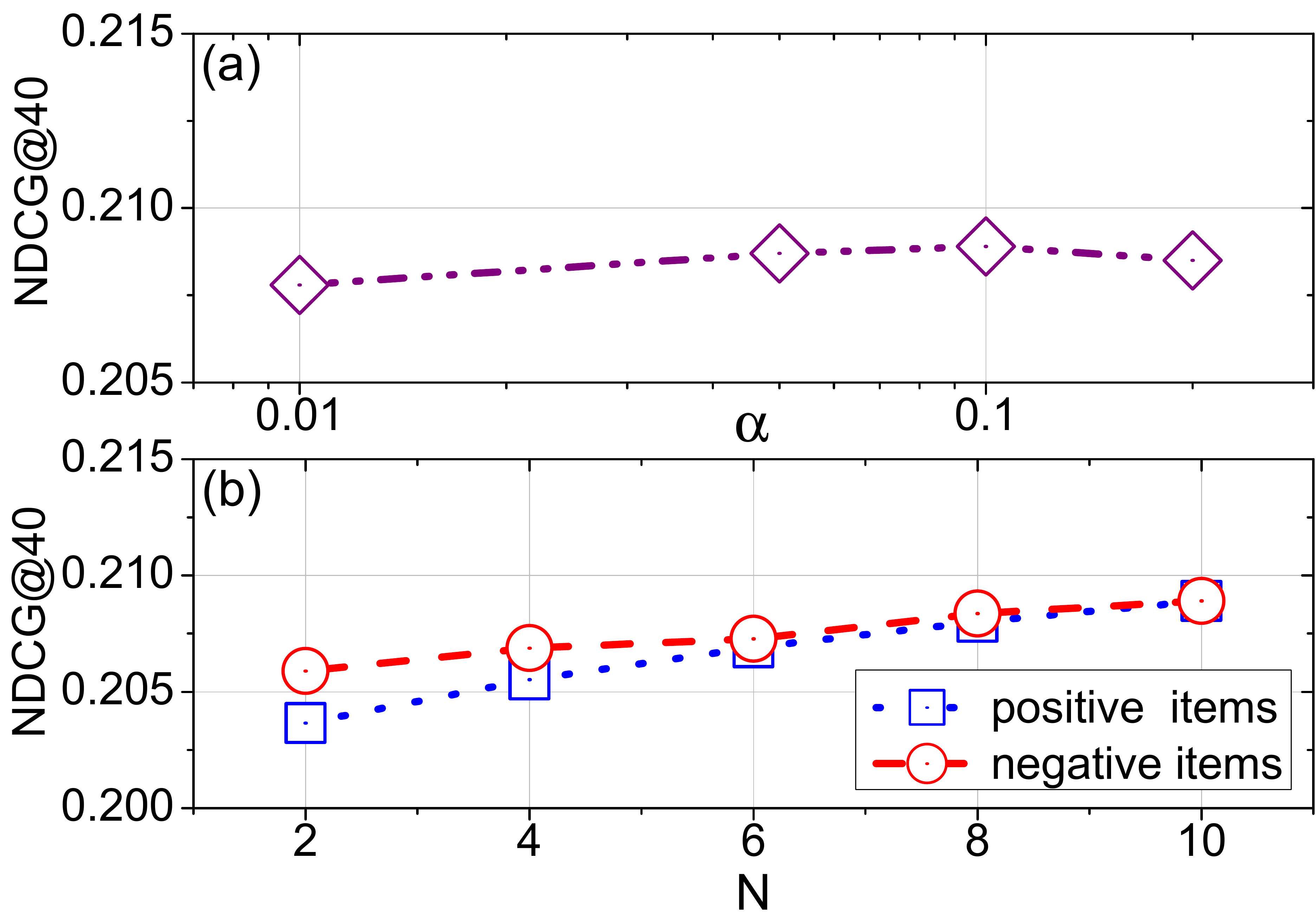}
	\caption{Parameter sensitiveness. (a) $\alpha$ that controls the pairwise regularization term. (b) $N$ that controls the length of state.}
	\label{fig:parameters}
\end{figure}

Figure \ref{fig:parameters} (a) shows the parameter sensitivity of $\alpha$ in offline recommendation task. The performance for recommendation achieves the peak when $\alpha = 0.1$. In other words, the the pairwise regularization term indeed improves the performance of the model; however, it is less important than the first term $\big(y-Q(s_{+}, s_{-}, a;\theta)\big)^{2}$ in Equation \ref{equ:L1_1}.

Figure \ref{fig:parameters} (b) demonstrates the parameter sensitivity of $N$ in offline recommendation task. We find that with the increase of $N$, the performance improves, and the performance is more sensitive with positive items. In other words, user's decision mainly depends on the items she/he clicked or ordered, but the skipped items also influence the decision making process.

%% file: 5RelatedWork_Conclusion.tex
\section{Related Work}
\label{sec:related_work}
In this section, we briefly review works related to our study. In general, the related work can be mainly grouped into the following categories.

The first category related to this paper is traditional recommendation techniques. Recommender systems assist users by supplying a list of items that might interest users. Efforts have been made on offering meaningful recommendations to users. Collaborative filtering~\cite{linden2003amazon} is the most successful and the most widely used technique, which is based on the hypothesis that people often get the best recommendations from someone with similar tastes to themselves~\cite{breese1998empirical}. Another common approach is content-based filtering~\cite{mooney2000content}, which tries to recommend items with similar properties to those that a user ordered in the past. Knowledge-based systems~\cite{akerkar2010knowledge} recommend items based on specific domain knowledge about how certain item features meet users needs and preferences and how the item is useful for the user. Hybrid recommender systems are based on the combination of the above mentioned two or more types of techniques~\cite{burke2002hybrid}. The other topic closely related to this category is deep learning based recommender system, which is able to effectively capture the non-linear and non-trivial user-item relationships, and enables the codification of more complex abstractions as data representations in the higher layers~\cite{zhang2017deep}. For instance, Nguyen et al.~\cite{nguyen2017personalized} proposed a personalized tag recommender system based on CNN. It utilizes constitutional and max-pooling layer to get visual features from patches of images. Wu et al.~\cite{wu2016personal} designed a session-based recommendation model for real-world e-commerce website. It utilizes the basic RNN to predict what user will buy next based on the click histories. This method helps balance the trade off between computation costs and prediction accuracy.

The second category is about reinforcement learning for recommendations, which is different with the traditional item recommendations. In this paper, we consider the recommending procedure as sequential interactions between users and recommender agent; and leverage reinforcement learning to automatically learn the optimal recommendation strategies. Indeed, reinforcement learning have been widely examined in recommendation field. The MDP-Based CF model can be viewed as approximating a partial observable MDP (POMDP) by using a finite rather than unbounded window of past history to define the current state~\cite{shani2005mdp}. To reduce the high computational and representational complexity of POMDP, three strategies have been developed: value function approximation~\cite{hauskrecht1997incremental}, policy based optimization~\cite{ng2000pegasus,poupart2005vdcbpi}, and stochastic sampling~\cite{kearns2002sparse}. Furthermore, Mahmood et al.~\cite{mahmood2009improving} adopted the reinforcement learning technique to observe the responses of users in a conversational recommender, with the aim to maximize a numerical cumulative reward function modeling the benefit that users get from each recommendation session. Taghipour et al.~\cite{taghipour2007usage,taghipour2008hybrid} modeled web page recommendation as a Q-Learning problem and learned to make recommendations from web usage data as the actions rather than discovering explicit patterns from the data. The system inherits the intrinsic characteristic of reinforcement learning which is in a constant learning process. Sunehag et al.~\cite{sunehag2015deep} introduced agents that successfully address sequential decision problems with high-dimensional combinatorial state and action spaces. Zhao et al.~\cite{zhao2017deep,zhao2018deep} propose a novel page-wise recommendation framework based on deep reinforcement learning, which can optimize a page of items with proper display based on real-time feedback from users.

\section{Conclusion}
\label{sec:conclusion}
In this paper, we propose a novel framework DEERS, which models the recommendation session as a Markov Decision Process and leverages Reinforcement Learning to automatically learn the optimal recommendation strategies. Reinforcement learning based recommender systems have two advantages: (1) they can continuously update strategies during the interactions, and (2) they are able to learn a strategy that maximizes the long-term cumulative reward from users. Different from previous work, we leverage Deep Q-Network and integrate skipped items (negative feedback) into reinforcement learning based recommendation strategies. Note that we design a novel architecture of DQN to capture both positive and negative feedback simultaneously. We evaluate our framework with extensive experiments based on data from a real e-commerce site. The results show that (1) our framework can significantly improve the recommendation performance; and (2) skipped items (negative feedback) can assist item recommendation.

There are several interesting research directions. First, in addition to positional order of items we used in this work, we would like to investigate more orders like temporal order. Second, we would like to validate with more agent-user interaction patterns, e.g., storing items into shopping cart, and investigate how to model them mathematically for recommendations. Finally, the items skipped by users may not be caused by users disliking them, but just not preferring as more as the items clicked/ordered or not viewing them in details at all. The week/wrong negative feedback may not improve or even reduce the performance when we consider the negative feedback. To capture stronger negative feedback, more information like dwell time can be recorded in users' behavior log and used in our framework.

%% file: 0KDD2018.bbl

\begin{thebibliography}{34}

%%% the Web address.

\ifx \showCODEN    \undefined \def \showCODEN     #1{\unskip}     \fi
\ifx \showDOI      \undefined \def \showDOI       #1{#1}\fi
\ifx \showISBNx    \undefined \def \showISBNx     #1{\unskip}     \fi
\ifx \showISBNxiii \undefined \def \showISBNxiii  #1{\unskip}     \fi
\ifx \showISSN     \undefined \def \showISSN      #1{\unskip}     \fi
\ifx \showLCCN     \undefined \def \showLCCN      #1{\unskip}     \fi
\ifx \shownote     \undefined \def \shownote      #1{#1}          \fi
\ifx \showarticletitle \undefined \def \showarticletitle #1{#1}   \fi
\ifx \showURL      \undefined \def \showURL       {\relax}        \fi
% invisible to TeX
\providecommand\bibfield[2]{#2}
\providecommand\bibinfo[2]{#2}
\providecommand\natexlab[1]{#1}
\providecommand\showeprint[2][]{arXiv:#2}

\bibitem[\protect\citeauthoryear{Akerkar and Sajja}{Akerkar and Sajja}{2010}]%
        {akerkar2010knowledge}
\bibfield{author}{\bibinfo{person}{Rajendra Akerkar} {and}
  \bibinfo{person}{Priti Sajja}.} \bibinfo{year}{2010}\natexlab{}.
\newblock \bibinfo{booktitle}{\emph{Knowledge-based systems}}.
\newblock \bibinfo{publisher}{Jones \& Bartlett Publishers}.
\newblock


\bibitem[\protect\citeauthoryear{Bellman}{Bellman}{2013}]%
        {bellman2013dynamic}
\bibfield{author}{\bibinfo{person}{Richard Bellman}.}
  \bibinfo{year}{2013}\natexlab{}.
\newblock \bibinfo{booktitle}{\emph{Dynamic programming}}.
\newblock \bibinfo{publisher}{Courier Corporation}.
\newblock


\bibitem[\protect\citeauthoryear{Breese, Heckerman, and Kadie}{Breese
  et~al\mbox{.}}{1998}]%
        {breese1998empirical}
\bibfield{author}{\bibinfo{person}{John~S Breese}, \bibinfo{person}{David
  Heckerman}, {and} \bibinfo{person}{Carl Kadie}.}
  \bibinfo{year}{1998}\natexlab{}.
\newblock \showarticletitle{Empirical analysis of predictive algorithms for
  collaborative filtering}. In \bibinfo{booktitle}{\emph{Proceedings of the
  Fourteenth conference on Uncertainty in artificial intelligence}}. Morgan
  Kaufmann Publishers Inc., \bibinfo{pages}{43--52}.
\newblock


\bibitem[\protect\citeauthoryear{Burke}{Burke}{2002}]%
        {burke2002hybrid}
\bibfield{author}{\bibinfo{person}{Robin Burke}.}
  \bibinfo{year}{2002}\natexlab{}.
\newblock \showarticletitle{Hybrid recommender systems: Survey and
  experiments}.
\newblock \bibinfo{journal}{\emph{User modeling and user-adapted interaction}}
  \bibinfo{volume}{12}, \bibinfo{number}{4} (\bibinfo{year}{2002}),
  \bibinfo{pages}{331--370}.
\newblock


\bibitem[\protect\citeauthoryear{Degris, White, and Sutton}{Degris
  et~al\mbox{.}}{2012}]%
        {degris2012off}
\bibfield{author}{\bibinfo{person}{Thomas Degris}, \bibinfo{person}{Martha
  White}, {and} \bibinfo{person}{Richard~S Sutton}.}
  \bibinfo{year}{2012}\natexlab{}.
\newblock \showarticletitle{Off-policy actor-critic}.
\newblock \bibinfo{journal}{\emph{arXiv preprint arXiv:1205.4839}}
  (\bibinfo{year}{2012}).
\newblock


\bibitem[\protect\citeauthoryear{Dupret and Piwowarski}{Dupret and
  Piwowarski}{2008}]%
        {dupret2008user}
\bibfield{author}{\bibinfo{person}{Georges~E Dupret} {and}
  \bibinfo{person}{Benjamin Piwowarski}.} \bibinfo{year}{2008}\natexlab{}.
\newblock \showarticletitle{A user browsing model to predict search engine
  click data from past observations.}. In \bibinfo{booktitle}{\emph{Proceedings
  of the 31st annual international ACM SIGIR conference on Research and
  development in information retrieval}}. ACM, \bibinfo{pages}{331--338}.
\newblock


\bibitem[\protect\citeauthoryear{Hauskrecht}{Hauskrecht}{1997}]%
        {hauskrecht1997incremental}
\bibfield{author}{\bibinfo{person}{Milos Hauskrecht}.}
  \bibinfo{year}{1997}\natexlab{}.
\newblock \showarticletitle{Incremental methods for computing bounds in
  partially observable Markov decision processes}. In
  \bibinfo{booktitle}{\emph{AAAI/IAAI}}. \bibinfo{pages}{734--739}.
\newblock


\bibitem[\protect\citeauthoryear{Hidasi, Karatzoglou, Baltrunas, and
  Tikk}{Hidasi et~al\mbox{.}}{2015}]%
        {hidasi2015session}
\bibfield{author}{\bibinfo{person}{Bal{\'a}zs Hidasi},
  \bibinfo{person}{Alexandros Karatzoglou}, \bibinfo{person}{Linas Baltrunas},
  {and} \bibinfo{person}{Domonkos Tikk}.} \bibinfo{year}{2015}\natexlab{}.
\newblock \showarticletitle{Session-based recommendations with recurrent neural
  networks}.
\newblock \bibinfo{journal}{\emph{arXiv preprint arXiv:1511.06939}}
  (\bibinfo{year}{2015}).
\newblock


\bibitem[\protect\citeauthoryear{J{\"a}rvelin and
  Kek{\"a}l{\"a}inen}{J{\"a}rvelin and Kek{\"a}l{\"a}inen}{2002}]%
        {jarvelin2002cumulated}
\bibfield{author}{\bibinfo{person}{Kalervo J{\"a}rvelin} {and}
  \bibinfo{person}{Jaana Kek{\"a}l{\"a}inen}.} \bibinfo{year}{2002}\natexlab{}.
\newblock \showarticletitle{Cumulated gain-based evaluation of IR techniques}.
\newblock \bibinfo{journal}{\emph{ACM Transactions on Information Systems
  (TOIS)}} \bibinfo{volume}{20}, \bibinfo{number}{4} (\bibinfo{year}{2002}),
  \bibinfo{pages}{422--446}.
\newblock


\bibitem[\protect\citeauthoryear{Kearns, Mansour, and Ng}{Kearns
  et~al\mbox{.}}{2002}]%
        {kearns2002sparse}
\bibfield{author}{\bibinfo{person}{Michael Kearns}, \bibinfo{person}{Yishay
  Mansour}, {and} \bibinfo{person}{Andrew~Y Ng}.}
  \bibinfo{year}{2002}\natexlab{}.
\newblock \showarticletitle{A sparse sampling algorithm for near-optimal
  planning in large Markov decision processes}.
\newblock \bibinfo{journal}{\emph{Machine learning}} \bibinfo{volume}{49},
  \bibinfo{number}{2} (\bibinfo{year}{2002}), \bibinfo{pages}{193--208}.
\newblock


\bibitem[\protect\citeauthoryear{Levy and Goldberg}{Levy and Goldberg}{2014}]%
        {levy2014neural}
\bibfield{author}{\bibinfo{person}{Omer Levy} {and} \bibinfo{person}{Yoav
  Goldberg}.} \bibinfo{year}{2014}\natexlab{}.
\newblock \showarticletitle{Neural word embedding as implicit matrix
  factorization}. In \bibinfo{booktitle}{\emph{Advances in neural information
  processing systems}}. \bibinfo{pages}{2177--2185}.
\newblock


\bibitem[\protect\citeauthoryear{Lin}{Lin}{1993}]%
        {lin1993reinforcement}
\bibfield{author}{\bibinfo{person}{Long-Ji Lin}.}
  \bibinfo{year}{1993}\natexlab{}.
\newblock \bibinfo{booktitle}{\emph{Reinforcement learning for robots using
  neural networks}}.
\newblock \bibinfo{type}{{T}echnical {R}eport}.
  \bibinfo{institution}{Carnegie-Mellon Univ Pittsburgh PA School of Computer
  Science}.
\newblock


\bibitem[\protect\citeauthoryear{Linden, Smith, and York}{Linden
  et~al\mbox{.}}{2003}]%
        {linden2003amazon}
\bibfield{author}{\bibinfo{person}{Greg Linden}, \bibinfo{person}{Brent Smith},
  {and} \bibinfo{person}{Jeremy York}.} \bibinfo{year}{2003}\natexlab{}.
\newblock \showarticletitle{Amazon. com recommendations: Item-to-item
  collaborative filtering}.
\newblock \bibinfo{journal}{\emph{IEEE Internet computing}}
  \bibinfo{volume}{7}, \bibinfo{number}{1} (\bibinfo{year}{2003}),
  \bibinfo{pages}{76--80}.
\newblock


\bibitem[\protect\citeauthoryear{Mahmood and Ricci}{Mahmood and Ricci}{2007}]%
        {mahmood2007learning}
\bibfield{author}{\bibinfo{person}{Tariq Mahmood} {and}
  \bibinfo{person}{Francesco Ricci}.} \bibinfo{year}{2007}\natexlab{}.
\newblock \showarticletitle{Learning and adaptivity in interactive recommender
  systems}. In \bibinfo{booktitle}{\emph{Proceedings of the ninth international
  conference on Electronic commerce}}. ACM, \bibinfo{pages}{75--84}.
\newblock


\bibitem[\protect\citeauthoryear{Mahmood and Ricci}{Mahmood and Ricci}{2009}]%
        {mahmood2009improving}
\bibfield{author}{\bibinfo{person}{Tariq Mahmood} {and}
  \bibinfo{person}{Francesco Ricci}.} \bibinfo{year}{2009}\natexlab{}.
\newblock \showarticletitle{Improving recommender systems with adaptive
  conversational strategies}. In \bibinfo{booktitle}{\emph{Proceedings of the
  20th ACM conference on Hypertext and hypermedia}}. ACM,
  \bibinfo{pages}{73--82}.
\newblock


\bibitem[\protect\citeauthoryear{Mnih, Kavukcuoglu, Silver, Graves, Antonoglou,
  Wierstra, and Riedmiller}{Mnih et~al\mbox{.}}{2013}]%
        {mnih2013playing}
\bibfield{author}{\bibinfo{person}{Volodymyr Mnih}, \bibinfo{person}{Koray
  Kavukcuoglu}, \bibinfo{person}{David Silver}, \bibinfo{person}{Alex Graves},
  \bibinfo{person}{Ioannis Antonoglou}, \bibinfo{person}{Daan Wierstra}, {and}
  \bibinfo{person}{Martin Riedmiller}.} \bibinfo{year}{2013}\natexlab{}.
\newblock \showarticletitle{Playing atari with deep reinforcement learning}.
\newblock \bibinfo{journal}{\emph{arXiv preprint arXiv:1312.5602}}
  (\bibinfo{year}{2013}).
\newblock


\bibitem[\protect\citeauthoryear{Mooney and Roy}{Mooney and Roy}{2000}]%
        {mooney2000content}
\bibfield{author}{\bibinfo{person}{Raymond~J Mooney} {and}
  \bibinfo{person}{Loriene Roy}.} \bibinfo{year}{2000}\natexlab{}.
\newblock \showarticletitle{Content-based book recommending using learning for
  text categorization}. In \bibinfo{booktitle}{\emph{Proceedings of the fifth
  ACM conference on Digital libraries}}. ACM, \bibinfo{pages}{195--204}.
\newblock


\bibitem[\protect\citeauthoryear{Moore and Atkeson}{Moore and Atkeson}{1993}]%
        {moore1993prioritized}
\bibfield{author}{\bibinfo{person}{Andrew~W Moore} {and}
  \bibinfo{person}{Christopher~G Atkeson}.} \bibinfo{year}{1993}\natexlab{}.
\newblock \showarticletitle{Prioritized sweeping: Reinforcement learning with
  less data and less time}.
\newblock \bibinfo{journal}{\emph{Machine learning}} \bibinfo{volume}{13},
  \bibinfo{number}{1} (\bibinfo{year}{1993}), \bibinfo{pages}{103--130}.
\newblock


\bibitem[\protect\citeauthoryear{Ng and Jordan}{Ng and Jordan}{2000}]%
        {ng2000pegasus}
\bibfield{author}{\bibinfo{person}{Andrew~Y Ng} {and} \bibinfo{person}{Michael
  Jordan}.} \bibinfo{year}{2000}\natexlab{}.
\newblock \showarticletitle{PEGASUS: A policy search method for large MDPs and
  POMDPs}. In \bibinfo{booktitle}{\emph{Proceedings of the Sixteenth conference
  on Uncertainty in artificial intelligence}}. Morgan Kaufmann Publishers Inc.,
  \bibinfo{pages}{406--415}.
\newblock


\bibitem[\protect\citeauthoryear{Nguyen, Wistuba, Grabocka, Drumond, and
  Schmidt-Thieme}{Nguyen et~al\mbox{.}}{2017}]%
        {nguyen2017personalized}
\bibfield{author}{\bibinfo{person}{Hanh~TH Nguyen}, \bibinfo{person}{Martin
  Wistuba}, \bibinfo{person}{Josif Grabocka}, \bibinfo{person}{Lucas~Rego
  Drumond}, {and} \bibinfo{person}{Lars Schmidt-Thieme}.}
  \bibinfo{year}{2017}\natexlab{}.
\newblock \showarticletitle{Personalized Deep Learning for Tag Recommendation}.
  In \bibinfo{booktitle}{\emph{Pacific-Asia Conference on Knowledge Discovery
  and Data Mining}}. Springer, \bibinfo{pages}{186--197}.
\newblock


\bibitem[\protect\citeauthoryear{Poupart and Boutilier}{Poupart and
  Boutilier}{2005}]%
        {poupart2005vdcbpi}
\bibfield{author}{\bibinfo{person}{Pascal Poupart} {and} \bibinfo{person}{Craig
  Boutilier}.} \bibinfo{year}{2005}\natexlab{}.
\newblock \showarticletitle{VDCBPI: an approximate scalable algorithm for large
  POMDPs}. In \bibinfo{booktitle}{\emph{Advances in Neural Information
  Processing Systems}}. \bibinfo{pages}{1081--1088}.
\newblock


\bibitem[\protect\citeauthoryear{Rendle}{Rendle}{2010}]%
        {rendle2010factorization}
\bibfield{author}{\bibinfo{person}{Steffen Rendle}.}
  \bibinfo{year}{2010}\natexlab{}.
\newblock \showarticletitle{Factorization machines}. In
  \bibinfo{booktitle}{\emph{Data Mining (ICDM), 2010 IEEE 10th International
  Conference on}}. IEEE, \bibinfo{pages}{995--1000}.
\newblock


\bibitem[\protect\citeauthoryear{Resnick and Varian}{Resnick and
  Varian}{1997}]%
        {resnick1997recommender}
\bibfield{author}{\bibinfo{person}{Paul Resnick} {and} \bibinfo{person}{Hal~R
  Varian}.} \bibinfo{year}{1997}\natexlab{}.
\newblock \showarticletitle{Recommender systems}.
\newblock \bibinfo{journal}{\emph{Commun. ACM}} \bibinfo{volume}{40},
  \bibinfo{number}{3} (\bibinfo{year}{1997}), \bibinfo{pages}{56--58}.
\newblock


\bibitem[\protect\citeauthoryear{Ricci, Rokach, and Shapira}{Ricci
  et~al\mbox{.}}{2011}]%
        {ricci2011introduction}
\bibfield{author}{\bibinfo{person}{Francesco Ricci}, \bibinfo{person}{Lior
  Rokach}, {and} \bibinfo{person}{Bracha Shapira}.}
  \bibinfo{year}{2011}\natexlab{}.
\newblock \showarticletitle{Introduction to recommender systems handbook}.
\newblock In \bibinfo{booktitle}{\emph{Recommender systems handbook}}.
  \bibinfo{publisher}{Springer}, \bibinfo{pages}{1--35}.
\newblock


\bibitem[\protect\citeauthoryear{Shani, Heckerman, and Brafman}{Shani
  et~al\mbox{.}}{2005}]%
        {shani2005mdp}
\bibfield{author}{\bibinfo{person}{Guy Shani}, \bibinfo{person}{David
  Heckerman}, {and} \bibinfo{person}{Ronen~I Brafman}.}
  \bibinfo{year}{2005}\natexlab{}.
\newblock \showarticletitle{An MDP-based recommender system}.
\newblock \bibinfo{journal}{\emph{Journal of Machine Learning Research}}
  \bibinfo{volume}{6}, \bibinfo{number}{Sep} (\bibinfo{year}{2005}),
  \bibinfo{pages}{1265--1295}.
\newblock


\bibitem[\protect\citeauthoryear{Sunehag, Evans, Dulac-Arnold, Zwols, Visentin,
  and Coppin}{Sunehag et~al\mbox{.}}{2015}]%
        {sunehag2015deep}
\bibfield{author}{\bibinfo{person}{Peter Sunehag}, \bibinfo{person}{Richard
  Evans}, \bibinfo{person}{Gabriel Dulac-Arnold}, \bibinfo{person}{Yori Zwols},
  \bibinfo{person}{Daniel Visentin}, {and} \bibinfo{person}{Ben Coppin}.}
  \bibinfo{year}{2015}\natexlab{}.
\newblock \showarticletitle{Deep Reinforcement Learning with Attention for
  Slate Markov Decision Processes with High-Dimensional States and Actions}.
\newblock \bibinfo{journal}{\emph{arXiv preprint arXiv:1512.01124}}
  (\bibinfo{year}{2015}).
\newblock


\bibitem[\protect\citeauthoryear{Taghipour and Kardan}{Taghipour and
  Kardan}{2008}]%
        {taghipour2008hybrid}
\bibfield{author}{\bibinfo{person}{Nima Taghipour} {and} \bibinfo{person}{Ahmad
  Kardan}.} \bibinfo{year}{2008}\natexlab{}.
\newblock \showarticletitle{A hybrid web recommender system based on
  q-learning}. In \bibinfo{booktitle}{\emph{Proceedings of the 2008 ACM
  symposium on Applied computing}}. ACM, \bibinfo{pages}{1164--1168}.
\newblock


\bibitem[\protect\citeauthoryear{Taghipour, Kardan, and Ghidary}{Taghipour
  et~al\mbox{.}}{2007}]%
        {taghipour2007usage}
\bibfield{author}{\bibinfo{person}{Nima Taghipour}, \bibinfo{person}{Ahmad
  Kardan}, {and} \bibinfo{person}{Saeed~Shiry Ghidary}.}
  \bibinfo{year}{2007}\natexlab{}.
\newblock \showarticletitle{Usage-based web recommendations: a reinforcement
  learning approach}. In \bibinfo{booktitle}{\emph{Proceedings of the 2007 ACM
  conference on Recommender systems}}. ACM, \bibinfo{pages}{113--120}.
\newblock


\bibitem[\protect\citeauthoryear{Turpin and Scholer}{Turpin and
  Scholer}{2006}]%
        {turpin2006user}
\bibfield{author}{\bibinfo{person}{Andrew Turpin} {and} \bibinfo{person}{Falk
  Scholer}.} \bibinfo{year}{2006}\natexlab{}.
\newblock \showarticletitle{User performance versus precision measures for
  simple search tasks}. In \bibinfo{booktitle}{\emph{Proceedings of the 29th
  annual international ACM SIGIR conference on Research and development in
  information retrieval}}. ACM, \bibinfo{pages}{11--18}.
\newblock


\bibitem[\protect\citeauthoryear{Wu, Wang, Hong, and Shi}{Wu
  et~al\mbox{.}}{2017}]%
        {wu2017returning}
\bibfield{author}{\bibinfo{person}{Qingyun Wu}, \bibinfo{person}{Hongning
  Wang}, \bibinfo{person}{Liangjie Hong}, {and} \bibinfo{person}{Yue Shi}.}
  \bibinfo{year}{2017}\natexlab{}.
\newblock \showarticletitle{Returning is Believing: Optimizing Long-term User
  Engagement in Recommender Systems}. In \bibinfo{booktitle}{\emph{Proceedings
  of the 2017 ACM on Conference on Information and Knowledge Management}}. ACM,
  \bibinfo{pages}{1927--1936}.
\newblock


\bibitem[\protect\citeauthoryear{Wu, Ren, Yu, Chen, Zhang, and Zhu}{Wu
  et~al\mbox{.}}{2016}]%
        {wu2016personal}
\bibfield{author}{\bibinfo{person}{Sai Wu}, \bibinfo{person}{Weichao Ren},
  \bibinfo{person}{Chengchao Yu}, \bibinfo{person}{Gang Chen},
  \bibinfo{person}{Dongxiang Zhang}, {and} \bibinfo{person}{Jingbo Zhu}.}
  \bibinfo{year}{2016}\natexlab{}.
\newblock \showarticletitle{Personal recommendation using deep recurrent neural
  networks in NetEase}. In \bibinfo{booktitle}{\emph{Data Engineering (ICDE),
  2016 IEEE 32nd International Conference on}}. IEEE,
  \bibinfo{pages}{1218--1229}.
\newblock


\bibitem[\protect\citeauthoryear{Zhang, Yao, and Sun}{Zhang
  et~al\mbox{.}}{2017}]%
        {zhang2017deep}
\bibfield{author}{\bibinfo{person}{Shuai Zhang}, \bibinfo{person}{Lina Yao},
  {and} \bibinfo{person}{Aixin Sun}.} \bibinfo{year}{2017}\natexlab{}.
\newblock \showarticletitle{Deep Learning based Recommender System: A Survey
  and New Perspectives}.
\newblock \bibinfo{journal}{\emph{arXiv preprint arXiv:1707.07435}}
  (\bibinfo{year}{2017}).
\newblock


\bibitem[\protect\citeauthoryear{Zhao, Xia, Zhang, Ding, Yin, and Tang}{Zhao
  et~al\mbox{.}}{2018}]%
        {zhao2018deep}
\bibfield{author}{\bibinfo{person}{Xiangyu Zhao}, \bibinfo{person}{Long Xia},
  \bibinfo{person}{Liang Zhang}, \bibinfo{person}{Zhuoye Ding},
  \bibinfo{person}{Dawei Yin}, {and} \bibinfo{person}{Jiliang Tang}.}
  \bibinfo{year}{2018}\natexlab{}.
\newblock \showarticletitle{Deep Reinforcement Learning for Page-wise
  Recommendations}.
\newblock \bibinfo{journal}{\emph{arXiv preprint arXiv:1805.02343}}
  (\bibinfo{year}{2018}).
\newblock


\bibitem[\protect\citeauthoryear{Zhao, Zhang, Ding, Yin, Zhao, and Tang}{Zhao
  et~al\mbox{.}}{2017}]%
        {zhao2017deep}
\bibfield{author}{\bibinfo{person}{Xiangyu Zhao}, \bibinfo{person}{Liang
  Zhang}, \bibinfo{person}{Zhuoye Ding}, \bibinfo{person}{Dawei Yin},
  \bibinfo{person}{Yihong Zhao}, {and} \bibinfo{person}{Jiliang Tang}.}
  \bibinfo{year}{2017}\natexlab{}.
\newblock \showarticletitle{Deep Reinforcement Learning for List-wise
  Recommendations}.
\newblock \bibinfo{journal}{\emph{arXiv preprint arXiv:1801.00209}}
  (\bibinfo{year}{2017}).
\newblock


\end{thebibliography}
